\documentclass[12pt,preprint]{aastex6}

\usepackage{epsfig}

\newcommand{\rs}{r_{_{\rm S}}}

\newcommand{\OmegaK}{\Omega_{\rm K}}

\newcommand{\Qjet}{P_{_{\rm jet}}}
\newcommand{\Qesc}{P_{\rm esc}}
\newcommand{\Qdisk}{P_{\rm disk}}

\newcommand{\rjet}{R_{_{\rm jet}}}
\newcommand{\rS}{R_{\rm S}}

\newcommand{\SgrA}{Sgr\,A$^*$}

\newcommand{\Mdot}{\dot{M}}
\newcommand{\rstar}{r_{_*}}
\def\N0{\dot N_0}

\newcommand{\begeq}{\begin{equation}}
\newcommand{\fineq}{\end{equation}}
\newcommand{\begfig}{\begin{figure}}
\newcommand{\finfig}{\end{figure}}
\newcommand{\begeqarray}{\begin{eqnarray}}
\newcommand{\fineqarray}{\end{eqnarray}}

\slugcomment{accpeted for publication in MNRAS}

\shorttitle{Jet Launching Radius in Low-Power AGNs}

\shortauthors{Le, Newman, \& Edge}

\begin{document}

\title{Jet Launching Radius in Low-Power Radio-Loud AGNs \\ In Advection-Dominated Accretion Flows}

\author{Truong Le, William Newman, and Brinkley Edge}

\affil{Department of Physics, Astronomy \& Geology, Berry College, Mount Berry, GA 30149, USA; tle@berry.edu}

\begin{abstract}
Using our theory for the production of relativistic outflows, we estimate 
the jet launching radius and the inferred mass accretion rate for 52 
low-power radio-loud AGNs based on the observed jet powers. Our analysis indicates 
that (1) a significant fraction of the accreted energy is required to convert the accreted 
mass to relativistic energy particles for the production of the jets near the event horizon, 
(2) the jets launching radius moves radially toward the horizon as the mass accretion 
rate or jets power increases, and (3) no jet/outflow formation is possible beyond $44$ 
gravitational radii.
\end{abstract}



\section{INTRODUCTION}

Astrophysical jets are highly collimated beams of high velocity outflows that originate from young stars, micro-quasars, gamma-ray bursts, and active galactic nuclei (AGNs). It is believed that low-power radio-loud AGNs harbor supermassive central black holes that contain hot, two-temperature advection dominated accretion flows (ADAFs) with significantly sub-Eddington accretion rates and powerful jets \citep[e.g.,][]{nkh97}, and this hot ADAF disks can efficiently accelerate the relativistic particles powering the jets \citep[e.g.,][hereafter LB05]{lb05}. It is believed that one efficient way to remove angular momentum from a disk is to eject it vertically into a jet or outflow, hence, suggesting that disk and jet are connected through accretion and ejection processes. The common understanding is that magnetohydrodynamics (MHD) are responsible for launching, accelerating, and collimating these jets, which involve the extraction of energy from the rotation (or spin) of the black hole or the accretion disk in order to power the jets/outflows~\citep[e.g.,][]{bz77,bp82,doe12}. However, the acceleration process due to the black hole spin still remains an open question~\citep[e.g.,][]{tak10,bs12,tch15,ino17}. For example, \citet{fgr10} found no single relation between the black hole spin and the jet power in X-ray binary jets, and \citet{nm12} pointed out that the effect of the black hole spin is not important when the compact jet is launched far away from the inner radius~\citep[$10-100 \, r_g$, where $r_g = GM/c^2$,][]{mnw05}. Moreover, \citet{md05} obtained an exact analytical solution to the Blandford-Znajek model, and showed that not only the energy extraction happens along the equatorial plane, but the energy is fed directly into the black hole.

Given the relative complexity of the electromagnetic models, LB05 first examined if the outflows can be understood through {\it first-order Fermi acceleration process} at a standing accretion isothermal shock, which parallels the early studies of cosmic-ray acceleration in supernova shock waves. Since then, we have studied and established that the acceleration of relativistic particles at a standing shock in ADAF inviscid and viscous disks can power the outflows frequently observed from low-power radio-loud AGNs (e.g., M87) and galactic black hole (e.g., \SgrA) candidates~\citep[e.g., LB05;][]{dbl09}. This model is depicted schematically in Figure~\ref{fig1}. In this scenario, the gas is accelerated gravitationally toward the central mass and the high-energy particles (the high-energy tail of the background Maxwellian distribution, which we call the test particles) experiences a shock transition due to an obstruction near the event horizon. These high-energy particles get accelerated at the shock location by first-order Fermi acceleration process and some become unbound and escape at the shock radius to form the jets/outflow, while others diffuse outward radially through the disk or advect across the event horizon into the black hole. For simplicity, we assume the escape of the relativistic particles occurs only at the shock location ($\rjet=r_*$), where the acceleration process is strongest. This allows us to model the escape of particles and energy without altering the global conservation equations. Here, the dynamical effect of the loss of energy is introduced through the shock jump conditions. Once the disk flow structure is determined, the transport equation for the relativistic number and energy densities and the particle distribution $f$ can be obtained self-consistently~\citep[e.g.,][]{lb07,bdl11}. The self-consistency of the calculation is established when the particle energy escape rate $\Qesc$ from the transport equation is equal to the energy lost rate from the disk $\Qdisk$, and these quantities must be equal to the observed estimated jets power $\Qjet$ of a particular source. The connection between the dynamical and the transport equations to the observed estimated jets power allows us to constrain the constant mass accretion rate $\dot{M}$ along the flow toward the event horizon and the jet launching radius $\rjet$ of an observed AGN. 

The existence of a shock is due to the presence of a ``centrifugal barrier'' situated close to the event horizon, and the post-shock sound speed and the disk thickness remain the same as the pre-shock values in an isothermal shock model. Furthermore, in the isothermal shock case, the shock must radiate away both energy and entropy at the shock location through the surface of the disk, and the energy lost from the shock can be identified to power the jet. From our studies, we showed that shock acceleration in a disk naturally produces a power-law energy distribution for the accelerated particles with a dominate power-law index of $\sim 4$ that has an equivalent supernova-driven shock of $\sim 6$ using the disk-shocked compression ratio. Our results indicated that the presence of a shock in a disk is very efficient in accelerating the relativistic particles and is similar to the cosmic-ray acceleration case~\citep[e.g.,][]{bo78}. Moreover, we have also shown that pressure of the accelerated relativistic particles can actually exceed the pressure of the thermal background gas in the vicinity of the shock~\citep[e.g., LB05;][]{bdl11}; and when this occurs, a ``two-fluid'' version of our model that includes the particle pressure, in analogy with the ``cosmic-ray modified shock'' scenario for cosmic-ray acceleration, must be included~\citep[e.g.,][]{bk01,dv81}.  Recently, \citet{leeb17} have included the ``two-fluid'' model for the structure of ADAF dics that properly accounts for the dynamical effect of the relativistic particle pressure. They conclude that the escape particles are mildly relativistic in the vicinity of a shock, and a smooth (shock-free) solutions cannot occur in diffusive, two-fluid discs.
\begfig[t] \hskip+0.2in \epsscale{0.6} \plotone{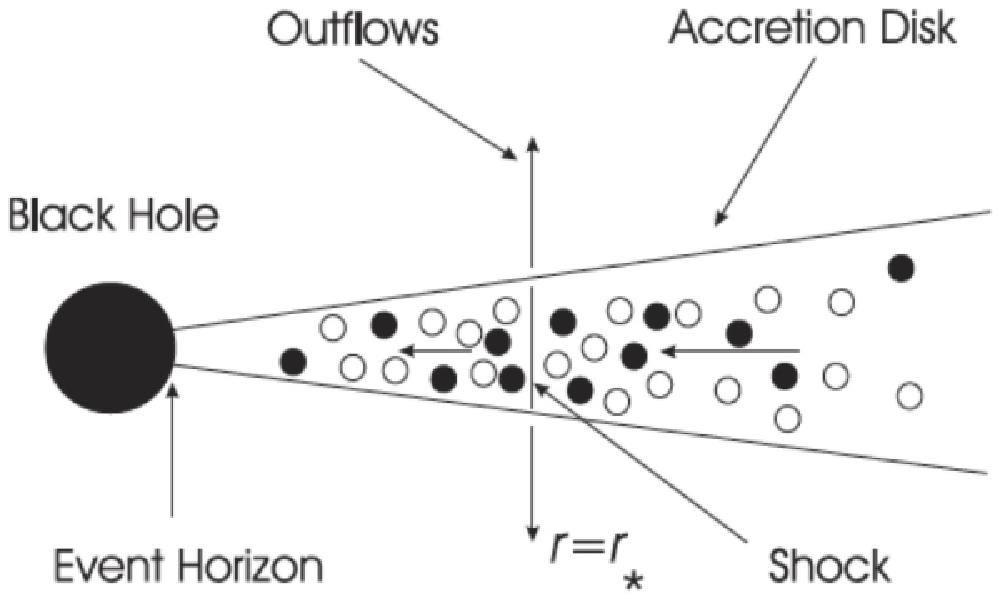}
\caption{\footnotesize Schematic representation of our disk-shock and disk-jet models. The schematic was taken from \citet{lb07} paper. The filled circles and the unfilled circles in the disk represent the test particles (or the high-energy tail of the background Maxwellian distribution) and the MHD scattering centers moving with the background gas, respectively. The test particles are injected at the shock location and get accelerated at the shock location by first-order Fermi acceleration process. Some will become unbound and escape at the shock radius to form the jets/outflow, while others will diffuse outward radially through the disk or advect across the event horizon into the black hole.}
\label{fig1}%
\vskip-0.0in %
\finfig

It is hard to answer in general, whether or not the escaped particles will produce collimated jets, but it has been assumed and demonstrated that magnetic field plays an important factor in collimating jets~\citep[e.g.,][]{bz77,mku01,mck06,tch15}. However, \citet{bro11} have demonstrated that unmagnetized or hydrodynamics jets can be collimated according to the strength of the jet-cocoon interaction and the collimation shock at the base of the jets. In this hydrodynamics collimated jet model, the relativistic energy particles are injected into the surrounding medium. The jet then propagates by pushing the matter in front of it, leading to the formation of a forward shock and a reverse shock at the jet's head. Matter that enters the head through the shocks is heated and flows sideways, and this leads to the formation of a pressured cocoon around the jet. If the cocoon's pressure is sufficiently high, it collimates the jet. From this picture, theoretically, our escaped particles could represent the jets injected particles at the base of the hydrodynamics jets. 

Independently, we notice that \citet{cha99}, \citet{ck16}, \citet{kc17}, and \citet{vc17}  have also invoked a jet launching point in the vicinity of a shock. From our own study of particle acceleration, the acceleration is strongest at the shock location, and therefore, should produce the strongest outflows at that point. However, some of the accelerated relativistic particles will diffuse or advect away from the shock, and can still escape downstream from the shock. These authors went beyond our own model by modeling the dynamics of the jets particles after they become unbound and leave the disk. However, the connection between the gas thermal particles from the disk and the relativistic particles that escape to form the jets is missing. Regardless, this is still very interesting because our model can provide the connection between the disk and the jet of the escape particles at the base of the jets. Nevertheless, the focus of our previous papers and in this paper is on particle acceleration mechanism and not on the jets' collimation or dynamics.

Whether the relativistic jet is accelerated via the Blandford-Znajek process or first-order Fermi process, one important question that has not been addressed is the correlation among the observed jets power, the jets launching radius, and the mass accretion rate at the event horizon for low-power radio-loud AGNs. It is evidence that to resolve the jet launching point observationally for extragalactic jet source either from an accretion disk or at the center of a black hole requires telescope with high angular resolution \citep[e.g.,][]{doe12,had16,wal16}, hence, the observed estimated jets launching radius for the low-power radio-loud AGNs are limited. Thus, if we could determine the correlations among these parameters, then we can obtain the mass accretion rate or jet launching radius of any low-power radio-loud AGNs. Since our model accounts the above physical parameters, it is therefore interesting to see if any correlations can be established.

We have shown previously that a basic disk structure that contains viscosity is still transonic and needs to go through two sonic points in a shock solution before reaching the speed of light at the horizon. Hence, in this work, we continue to utilize an inviscid disk to estimate the jet launching radius, since we expect the general results remain the same for a viscous disk. Using 8 low-power radio-loud sources in~\citet[][hereafter A06]{all06}, we estimate the jet locations $\rjet$ and the inferred mass accretion rates $\dot{M}$ from our model based on the respective observed estimated jet powers $\Qjet$. We then explore potential correlation properties that could exist among these parameters. From these correlations, we estimate the mass accretion rates and the jets launching radii for M87, \SgrA, and  32 other sources from \citet[][hereafter BBC08]{bbc08}. Finally, we check to see if the shock locations from our model are stable to justify sources that contain episodic or continuous jets/outflows. The rest of the paper is organized as follows. In Section~2, we briefly discuss the equations, the assumptions that describe the dynamics of the disks and the connection to the outflows/jets, and the physical parameters that allow the model to constrain the mass accretion rate and the jets launching radius based on the observed estimated jet power. In Section~3, we apply our model to the selected sources and discuss the results of the correlations, and conclude our paper in Section~4.

\section{DYNAMICAL MODEL}

In this section, we give an overview of a general procedure that we developed in the LB05 paper to construct the dynamical profiles and the transport equation for the relativistic energy particles that allow us to estimate the jet launching radius $\rjet$ and the constant mass accretion $\dot{M}$ along a flow toward the event horizon. For completeness, we restate all relevant equations from that paper in Sections 2.1-2.3.

\subsection{Hydrodynamics Equations}
\citet{cha89} and \citet{ac90} investigated the structure of a one-dimensional, steady state, axisymmetric inviscid accretion flow based on the vertically average conservation of mass, radial momentum, angular momentum, and internal energy, which are described by
\begeq %
  \frac{1}{r} \frac{d}{dr} (r v \rho H) = 0, %
  \label{eq1} %
\fineq

\begeq %
  v \frac{dv}{dr} + %
  \frac{1}{\rho} \frac{dP}{dr} - r(\Omega^2 - \Omega_{\rm K}^2) = 0, %
\label{eq2}
\fineq

\begeq  %
\ell \equiv r^2 \Omega =  {\rm constant} \, , %
\label{eq3} %
\fineq

\begeq %
  v \frac{dU}{dr} - \frac{\gamma U}{\rho} \left(v \frac{d\rho}{dr} \right) = 0,  %
\label{eq4} %
\fineq
respectively, where $v$ is defined to be negative for the radial velocity inflow, $\rho$, $\Omega$, $\ell$, $H$, $\OmegaK$, and $U$ are the mass density, the angular velocity, the accreted specific angular momentum, the disk half-thickness, the Keplerian angular velocity, and the internal energy density, respectively, and finally, $P=(\gamma -1) U$ is the gas pressure. These quantities represent a vertical average over the disk structure and, the specific heats ratio, $\gamma$, is constant throughout the flow.  We employed the above conservation equations to make the connection between the disk structure and the acceleration of the relativistic particles to the outflows/jets.

The model utilized the pseudo-Newtonian gravitational potential per unit mass and is given by
\begeq  %
\Phi(r) = -\frac{G M}{r-\rs} \ , %
\label{eq5} %
\fineq
to give the effects of general relativity, where $\rs = 2 GM/c^2$ is the Schwarzschild radius for a black hole of mass $M$~\citep{pw80}.  Using the pseudo-Newtonian potential, the Keplerian angular velocity $\Omega_{\rm K}$ of matter in a circular orbit at radius $r$ is given as
\begeq %
\Omega_{\rm K}^2 = \frac{G M}{r(r-\rs)^2} = \frac{1}{r}\frac{d\Phi}{dr} \ . %
\label{eq6} %
\fineq
The disk half-thickness $H$ in Equation (\ref{eq1}) is given by the standard hydrostatic prescription
\begeq %
H = \frac{a}{\Omega_{\rm K}} \ ,  %
\label{eq7} %
\fineq
where $a$ represents the adiabatic sound speed
\begeq %
a \equiv \left(\frac{\gamma P}{\rho}\right)^{1/2} ,\ %
\label{eq8} %
\fineq
and the pressure and density are related according to
\begeq %
P = C_0 \rho^\gamma \ , %
\label{eq9} %
\fineq
where the constant parameter $C_0$ is closely related to entropy of gas that is conserved along a flow except at a shock location due to energy loss. In this work, the flow is adiabatic everywhere except at a shock front due to the energy loss to satisfy the isothermal property of gas.

\subsection{The Critical Points, Isothermal Shock Jump Conditions \& the Estimated Physical Quantities $\rjet$ and $\dot{M}$}

In a steady and inviscid ADAFs disks, the mass transport rate $\dot{M}$, the angular momentum transport rate $\dot{J}$, and the energy transport rate $\dot{E}$ in Equations (4), (5), and (6) from LB05, respectively, are conserved \citep[e.g.,][]{bl03,bs05}, and they can be rewritten as 
\begeq
  \dot{M} = -4 \pi r H \rho v \, ,
\label{eq10}
\fineq
\begeq
  \dot{J} = \dot{M} r^2 \Omega \, ,
\label{eq11}
\fineq
and
\begeq
  \dot{E} = \dot{M} \left(\frac{v^2_{\phi}}{2} + \frac{v^2}{2} + \frac{P + U}{\rho} + \Phi \right) \, ,
\label{eq12}
\fineq
where $\rho$, $H$, $v$, $P$ and $U$ are defined as above and $v_{\phi} = r \Omega$ is the azimuthal velocity. Under the inviscid flow assumption, the disk model depends on the conserved energy transport rate per unit mass $\epsilon \equiv \dot{E}/\dot{M}$, the conserved angular momentum transport per unit mass $\ell \equiv \dot{J}/\dot{M}$ (cf. Equation (\ref{eq3})), and $\gamma$ as discussed above. The value of $\epsilon$ will jump at the location of an isothermal shock if one is present, otherwise, it will remain constant since there are no radiative losses. However, the value of $\ell$ remains constant since the flow is inviscid. From Equations (13) and (15) in LB05, except at the shock location, we have the conserved accreted energy transport rate
\begeq %
  \epsilon \equiv {\ell^2 \over {2 r^2}} + {v^2 \over 2} + {a^2 \over \gamma-1} - {GM \over r-\rs} \, ,%
\label{eq13} %
\fineq
and the conserved ``entropy parameter''
\begeq %
  K \equiv -v \, a^{(\gamma+1)(\gamma-1)} \, r^{3/2} \, (r-\rs) \ , %
\label{eq14} %
\fineq
respectively, and $\epsilon > 0$ represents an inward flow of energy into the black hole. When a shock is present, we denote the subscripts ``-'' and ``+'' to quantities measured in pre-shock  and post-shock, respectively. In the isothermal shock model, $K$ and $\epsilon$ have smaller values in the post-shock region ($K_{+}$, $\epsilon_{+}$) compared with the pre-shock region ($K_{-}$, $\epsilon_{-}$).

The critical points in a flow are obtained by using Equation (19) in LB05 and are fully discussed in Section~3.3 of that paper. 
Four solutions for the critical radius can be obtained and they are denoted as $r_{c4}$, $r_{c3}$, $r_{c2}$, and $r_{c1}$ in order of increasing radius. The critical radius $r_{c4}$ lies inside the event horizon, but the other three are located outside the horizon and so they are physically relevant. There are three possible types of critical points, namely,  the O-type, X-type, and $\alpha$-type. The critical point $r_{c2}$ is unphysical because it is an O-type, where the values for the derivatives for $dv/dr$ are complex. The root $r_{c3}$ is an X-type critical point, where a shock-free solution always exists, and $r_{c1}$ (hereafter $r^{\rm out}_{c} = r_{c1}$ is the outer critical point) is an $\alpha$-type critical point.  However, any accretion flow that passes through an $\alpha$-type critical point must undergo a shock transition at $\rstar$~\citep{ac90}. After crossing the shock, the subsonic gas must pass through another ($\alpha$-type) critical point $r^{\rm in}_{c}$ (the inner critical point) to become supersonic before entering the black hole. A specific example of a shock flow solution is  discussed in Section~2.3.

The radius of the steady isothermal shock, denoted by $\rstar$, is determined along with the structure of the disk by satisfying the velocity and energy jump conditions (see Equations (41) and (44) in LB05),
\begeqarray %
  {\cal R}_{_{*}}^{-1} \equiv {v_+ \over v_-} = {1 \over \gamma \,
  {\cal M}_-^2} 
\label{eq17} \\ %
\nonumber \\ %
\Delta\epsilon \equiv
  \epsilon_+-\epsilon_- = {v_+^2 - v_-^2 \over 2} \ , %
\label{eq18} %
\fineqarray
where ${\cal M}_{-} \equiv - v_-/a_-$ is the upstream Mach number and ${\cal R}_{_{*}}$ is the shock compression ratio. For an isothermal shock, $\Delta \epsilon$ is negative because the energy transport rate $\epsilon$ drops from the upstream value $\epsilon_-$ to the downstream value $\epsilon_+$. The downstream flow must therefore pass through a new inner critical point located at $r^{\rm in}_{c}< \rstar$. This inner critical point and the flow structure in the post-shock region are computed using the downstream value $\epsilon_+$, while the outer critical point and the flow structure in the pre-shock region are computed using the upstream value $\epsilon_-$. Depending on the values of ($\epsilon_-$, $\ell$, and $\gamma$), as will be discussed in Section~2.3, one shock flow solution or two different  shock flow solutions that go through the same outer sonic point $r^{\rm out}_{c}$ can be obtained. Furthermore, the numerical steady-state solution for the inflow speed $v(r)$, gas density $\rho(r)$, and entropy parameter $K(r)$ must satisfy the isothermal shock jump conditions at $\rstar$ (see Equations (41), (42), and (43) in LB05).

In our model, the jet launching radius $\rjet$ is at the shock radius $ r_*$. The estimated values for $\rjet$ and $\Mdot$ of a low-power radio-loud source from our model is related to the observed jet power $\Qjet$ by
\begeq 
\Qdisk = \Qjet = \Qesc, 
\label{eq1b}
\fineq
where $\Qdisk = \dot M \triangle \epsilon \, \propto \rm erg \, s^{-1}$ is the energy lost rate from the disk, $\dot{M}$ is the constant mass accretion rate along the flow toward the horizon, and $\Delta\epsilon$ is the jump in the energy inflow rate. Here, $\Qesc$ is the calculated particle energy escape rate from the transport equation and is given by
\begeq
\Qesc = 4 \, \pi \, r_* \, H_* \, A_0 \, c \, U_* \propto {\rm erg \, s^{-1}}
\label{eq1c}
\fineq
with $H_*$, $U_*$, and $A_0$ being the disk half-height, the relativistic particle energy density, and the escape parameter at the shock location, respectively. Analysis of the three-dimensional random walk of the escaping particles yields 
\begeq 
A_0 = \left(3 \kappa_0 v_* \rS \over c \, H_*\right)^2\left(\frac{r_*}{\rS} -1 \right)^4 < 1 \ , 
\label{eq2a} 
\fineq
where $v_*$ is the mean flow speed at the shock, $\kappa_0$ is the diffusion coefficient, and $\rS$ is the Schwarzschild radius. Equations (\ref{eq1b}), (\ref{eq1c}), and (\ref{eq2a}) are from Equations (45), (94), and (71) in LB05, respectively. To obtain a self-consistent result, the particle energy escape rate $\Qesc$ at the shock location must equal to the total energy lost rate $\Mdot \Delta\epsilon$ from the dynamical model, and these values are required to be equal to the observed jet power $\Qjet$ for a given $\dot M$. This condition, however, is not automatically satisfied and thus it constrains $\epsilon_-$, $\epsilon_+$, $\kappa_0$, and $\dot M$. Once the conditions in Equations (\ref{eq1b}) are satisfied, the mass accretion rate $\dot{M}$ and the associated jet launching radius $\rjet$ are constrained. However, if none of the ($\epsilon_{_{-}}, \epsilon_{_{+}}, \ell$) and $\kappa_0$ parameters from the model for a given $\dot M$ produces the observed jet power $\Qjet$, then we modify $\dot{M}$ and search for a new set of ($\epsilon_{_{-}}, \epsilon_{_{+}}, \ell$) and $\kappa_0$ values that give a jet power consistent with observations.

\subsection{Parameter Space Configuration \& Steady-State Solutions}

In this section, we briefly discuss the construction of the steady-state solutions. The search for a flow solution for a specific disk-shocked system begins with the selection of the parameters ($\epsilon_{_{-}}, \ell, \gamma$). In earlier work and in this paper, we set $\gamma = 1.5$ to reflect the contributions to the pressure from the gas and the equipartition magnetic field \citep[e.g.,][]{nkh97}, hence, only $\epsilon_{_{-}}$ and $\ell$ remain to be determined. The wedge in Figure~(\ref{fig2}a) represents the ($\epsilon_{_{-}}$, $\ell$)-parameter space for $\gamma=1.5$, where shocks can form for an isothermal shock type (for the construction of the wedge, see Section~4.2 in LB05). For an isothermal shock model, the post-shock entropy value is larger than the pre-shock entropy value as discussed earlier. The data points below $\ell = 3.5$ are the parameters values for all nine sources from A06, and a data point at $\ell=3.75$ is used to demonstrate a stable shock.

It is important to mention here that for our model to be able to address sources that contains either episodic or continuous jets/outflows, the shock location must be stable. \citet{le16} utilized a linear perturbation calculation and examined the stability of a disk structure between the shock point and the inner critical point to determine if a shock location is stable or unstable. The results of the calculation provide different modes of oscillations, which include the Z-mode (zero frequency of oscillation), F-mode (fundamental frequency or first harmonic), and the overtones (higher harmonics).  The oscillation period of the perturbed shock wave is determined by $\delta_i$ and the growth (unstable) or damping (stable) rate by $\delta_r$. Hence, for a shocked disk structure to be stable, every modes need to have $\delta_r < 0$. Thus, Figure~(\ref{fig2}b) demonstrates a disk flow structure with a stable shock with a disk-parameter that lies in the stable region as indicated by a single dot at $\ell = 3.75$ in  Figure~(\ref{fig2}a). The white colored contours in Figure~(\ref{fig2}b) indicate that a specific frequency exists for a particular mode, and that a selected eigenvalues have satisfied the boundary conditions set in the problem, while a large blue colored region does not, for example. For clarity, in Figure (2b), each contour indicates different level of intensity of a selected eigenmodes ($\delta_r, \delta_i$) corresponding to having the perturbed inflow velocity goes to zero at the inner sonic boundary of the perturbation calculation. If a contour has a high intensity level, then this implies a selected eigenmodes is closely satisfied a perturbed boundary condition. The white colored contours indicate a region of high intensity. In this work, we are interested in knowing if a shock location is stable ($\delta_r < 0$) or unstable ($\delta_r > 0$), and not a specific value of the eigenfrequency ($\delta_i$). To search for a specific eigenfrequency of a particular mode, we would have to subdivide the white contour region of that mode to even smaller grids. In a case if a particular mode is unstable, our model indicates that a selected source should not contains episodic or continuous outflows at that particular frequency; it does not necessary imply that particular shock location is unstable. We will address more about this topic near the end of Section~3.
\begfig[t] \hskip-0.0in \epsscale{0.95} \plottwo{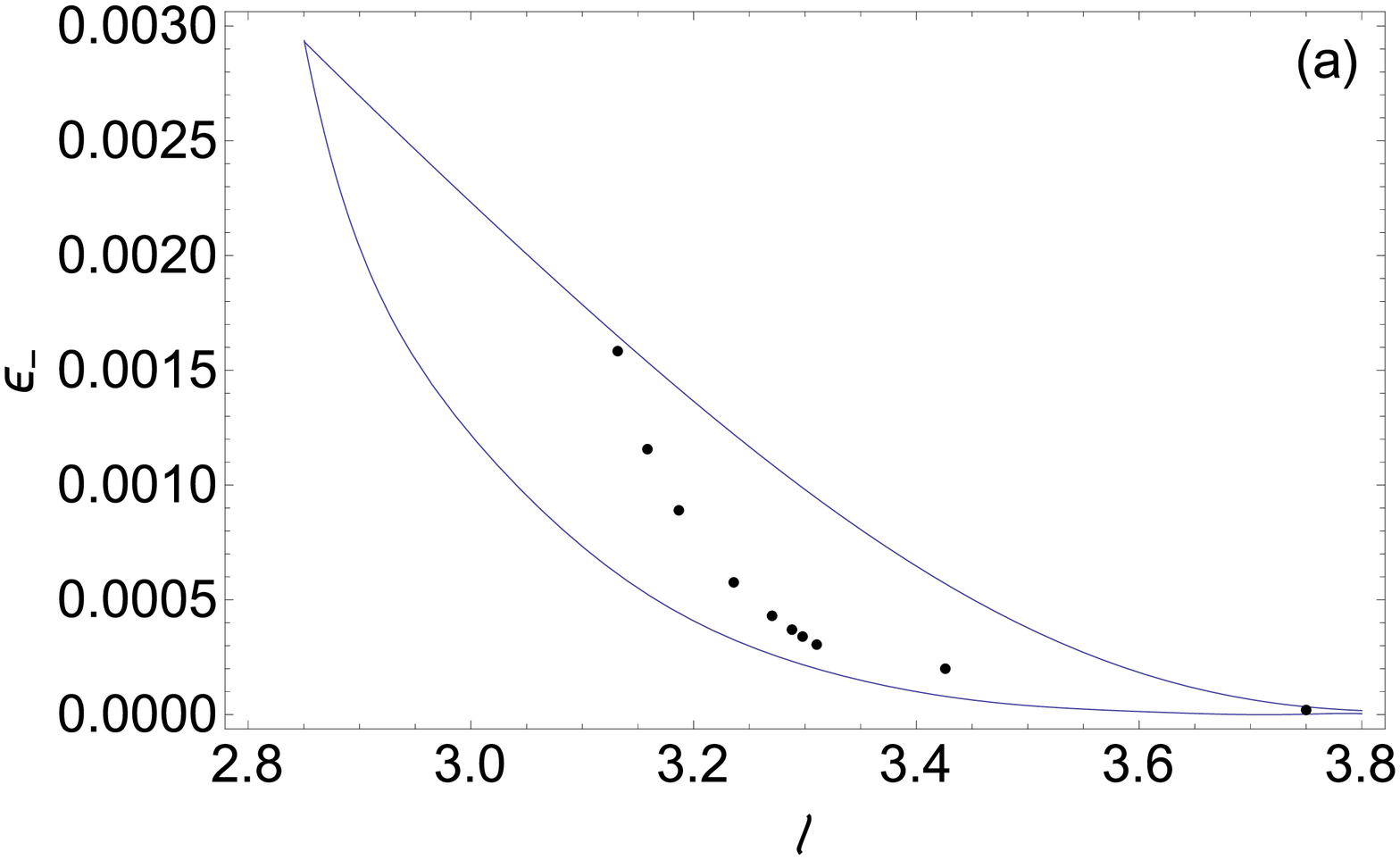}{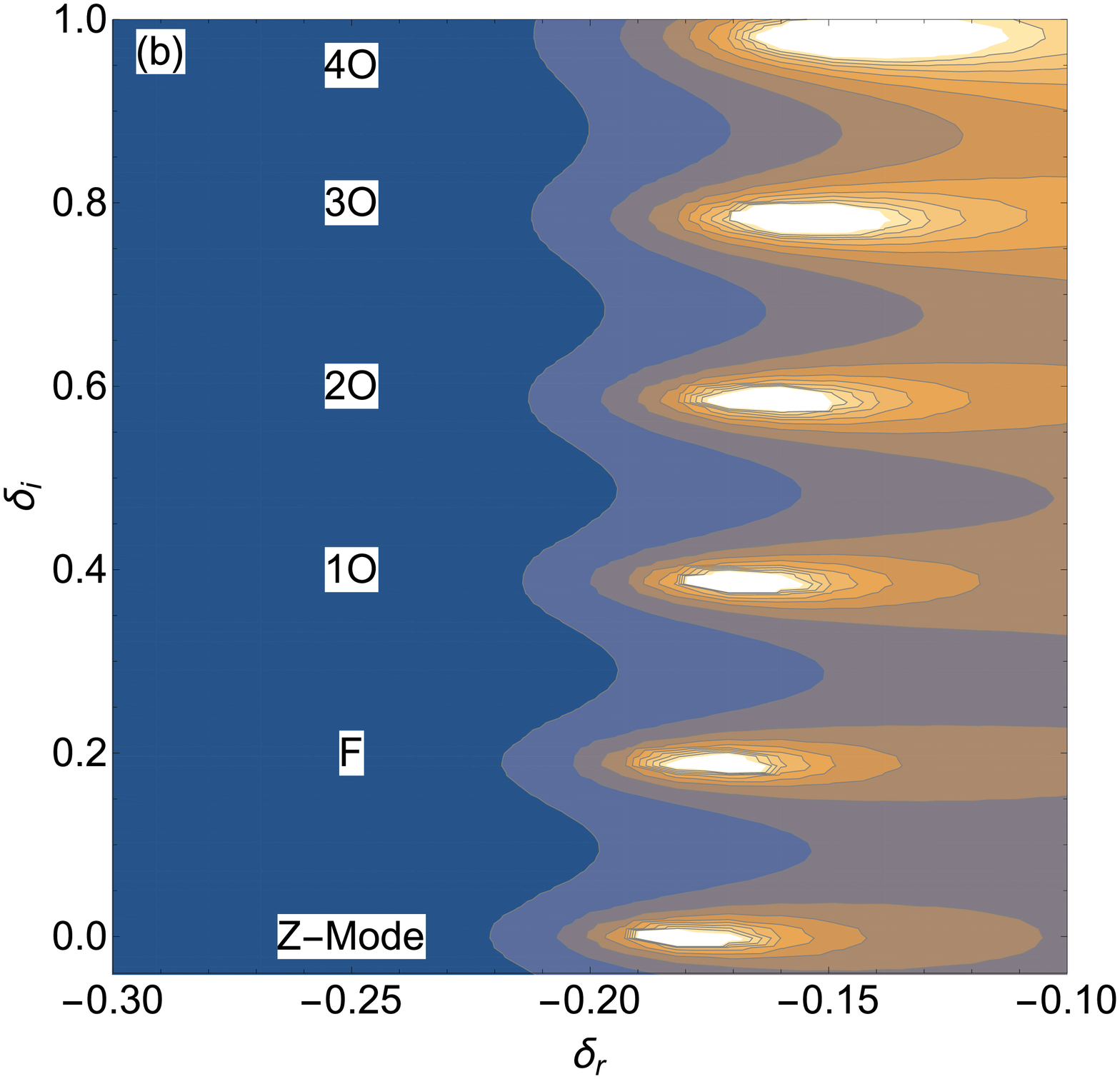}
\caption{\footnotesize ({\it a}) This plot represents the ($\epsilon_{-}, \ell$) parameter space for an inviscid ADAF disk with $\gamma = 1.5$,  where shocked and shock-free solutions are possible within the wedged region. The nine solid data points are the disks parameter values for the sources in A06. A single solid point at $\ell = 3.75$ is used to demonstrate a stable shock as illustrated in Figure (2b). ({\it b}) This plot illustrates a stable or unstable eigenmode of a disk structure. A white colored contour indicates that a specific frequency mode (e.g., the Z-mode, F-mode, and the overtones (1O, 2O, 3O, and 4O)) exists for the selected disk structure. Each contour indicates different level of intensity of a selected eigenmodes ($\delta_r, \delta_i$). The white colored contours indicate a region of high intensity. If a contour has a high intensity level, then this implies a selected eigenmodes is closely satisfied a perturbed boundary condition in the perturbation calculation. The oscillation period of the perturbed shock wave is determined by $\delta_i$ and the growth (unstable, $\delta_r > 0$) or damping (stable, $\delta_r < 0$) rate by $\delta_r$. This plot demonstrates a stable disk structure for ($\epsilon_{-} = 0.00002, \ell = 3.75$)-parameter values, where the Z-mode, F-mode, and the overtones are stable.}
\label{fig2} %
\finfig

For given values of ($\epsilon_{_{-}}, \epsilon_{_{+}}, \ell, \gamma$), a unique flow structure can be obtained using the procedures discussed above or in Sections~3 and 4 of LB05 that yields a numerical solution for $v(r)$ and $a(r)$ (e.g., Figure~\ref{fig3}a represents a general profile of the inflow speed and gas sound speed).  As discussed earlier, we set $\gamma=1.5$, and the acceptable values for $\epsilon_{_{-}}$, $\epsilon_{_{+}}$ and $\ell$ are constrained by the inferred estimated $\dot{M}$ and the observed estimated $\Qjet$ of a specific object. Figures~(\ref{fig3}a) and (\ref{fig3}b) presented here are for source NGC 507 to illustrate the use of our model, and the essential properties of these Figures and their model parameters are discussed below. We will discuss the initial guess value of $\dot M$ in Section~3. When a shock is present in the flow (see Figure~\ref{fig3}a), the gas passes through two critical points, at $r = r^{out}_c$ and $r =r^{in}_c$, where $r^{out}_c$ and $r^{in}_c$ are the outer and inner sonic points, respectively. Consequently, the isothermal shock radius $r_*$ must be determined self-consistently by satisfying the velocity and energy jump conditions as discussed in Section~2.2. Since the disk model that we consider here contains an isothermal shock, this implies that there will be a loss of energy at the shock radius due to the jump in energy $\triangle \epsilon = \epsilon_{+} - \epsilon_{-}$, where $\triangle \epsilon < 0$ implies a loss of energy as discussed in Equation (\ref{eq18}). As a result, the energy transport rate $\epsilon$ drops from the upstream value $\epsilon_-$ to the downstream value $\epsilon_+$ (see Fig.~\ref{fig3}b). 
\begfig[t] \hskip-0.0in \epsscale{1.1} \plottwo{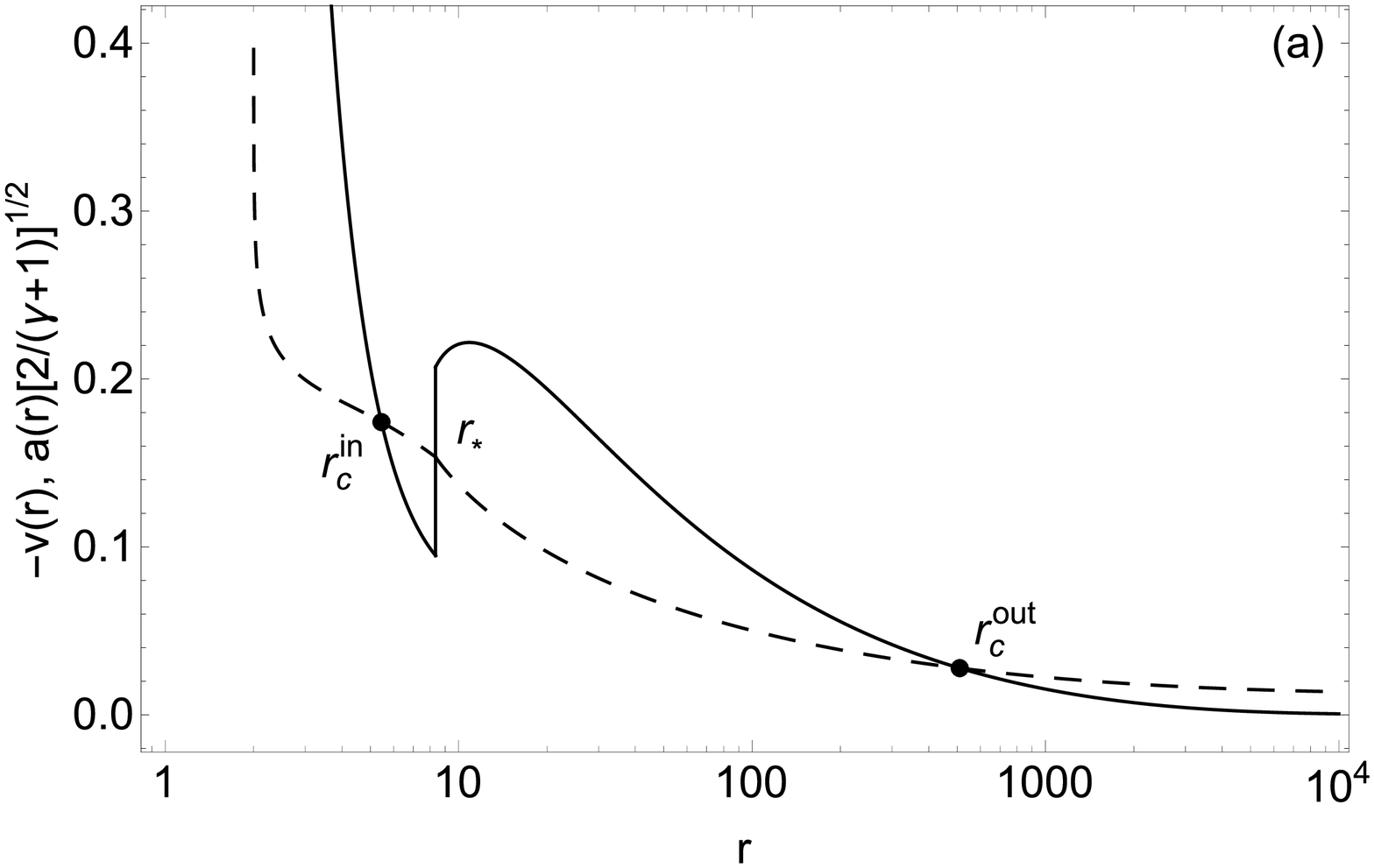}{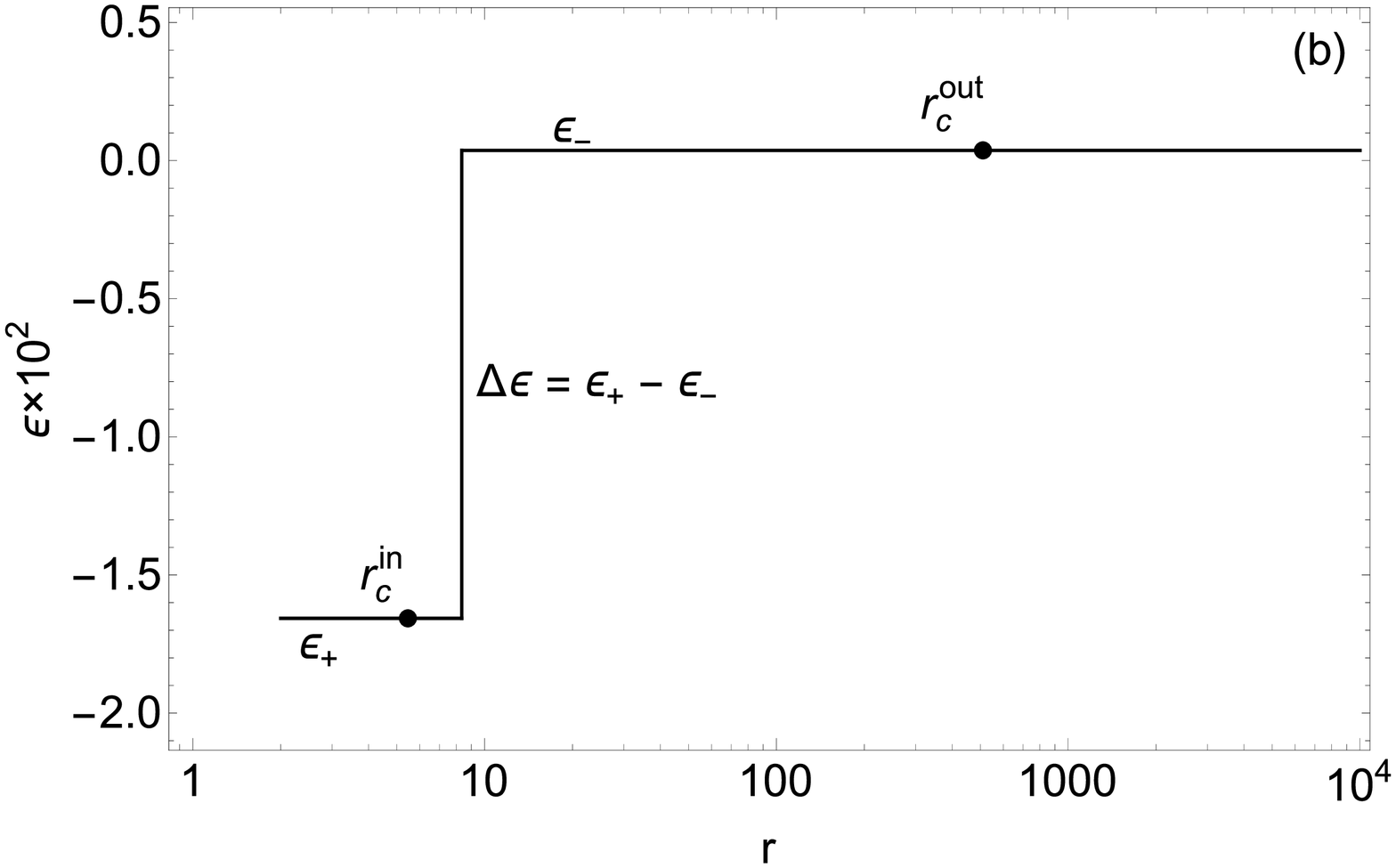} 
\vskip-0.1in
\caption{\footnotesize ({\it a}) This plot illustrates the gas inflow velocity $v$ (solid) and the isothermal sound speed $a$ 
(dashed) plotted as functions of the radius $r$ in units of $GM/c^2$. The flow is subsonic at large distance and becomes supersonic after crossing the outer sonic point $r^{out}_c$. The flow is shocked at the shock location $r_*$ and becomes subsonic. Before  entering the event horizon, it goes through the inner sonic point $r^{in}_c$ to reach supersonic speed. ({\it b}) This is the energy transport rate $\epsilon$ plotted as a function of radius $r$. Because this is an isothermal shock model, at the shock location, the accreted specific energy is dropped from the upstream $\epsilon_{-}$ to the downstream $\epsilon_{+}$ values. Both plots are model presentation for NGC 507.} 
\label{fig3} %
\finfig

\section{APPLICATIONS TO 52 LOW-POWER RADIO-LOUD AGNs}

\subsection{Model Parameter Selection}
In Table~\ref{tbl-1} the first four columns, respectively, are the source name, black hole mass, observed estimated jet power, and the estimated Bondi mass accretion rate from A06, while the fifth and sixth columns, respectively, are the estimated mass accretion rate and the jet launching location from this work. A06 estimate the total jet power as $P_{\rm jet} = P_0 (L_\nu/L_0)^{12/17}$ based on the measured radio-core luminosity $L_\nu$ with a constant $P_0 = 1.0^{+1.3}_{-0.6} \times 10^{44}$ erg s$^{-1}$ when $L_0$ is fixed at $L_0 = 7 \times 10^{29}$ erg Hz$^{-1}$ s$^{-1}$~\citep{hei07}. They also estimate the Bondi mass accretion rate using the Bondi approximation relation as $\dot{M}_{_{\rm B}} = \lambda  \pi r^2_{_{\rm B}} \rho_{_{\rm B}} a_{_{\rm B}}$, where  $\lambda$ is the normalized accretion rate coefficient that depends upon the adiabatic index of the accreting gas with $\lambda=0.25$ for an adiabatic index 5/3, and $\rho_{_{\rm B}}$ and $a_{_{\rm B}}$ are the gas density and sound speed, respectively, at the Bondi radius $r_{_{\rm B}}  = 2 G M_{\rm BH}/a^2_{_{\rm B}}$ with $M_{\rm BH}$ being the mass of the central black hole \citep[see][A06, for more details]{bon52}. The model parameters for each sources are in Table~\ref{tbl-2}, where from the first to the ninth columns, respectively, are the source name, specific accreted angular momentum $\ell$, upstream $\epsilon_{_{-}}$ and downstream $\epsilon_{_{+}}$ specific accreted energy, outer $r_{c}^{out}$ and inner $r_{c}^{in}$ sonic points, shock location $r_{*}$, upstream Mach number $\mathcal{M}_{-} $, and the compression ratio ${\cal R}_{_{*}}$ at the shock location. It is important to remind the reader here that the observed estimated Bondi mass accretion rate $\dot{M}_{_{\rm B}}$ obtained by A06 assumed the adiabatic index of the accreting gas with a value of $5/3$, while in our model we assume a value of $1.5$ for the accreting gas to reflect the contributions to the pressure from the gas and the equipartition magnetic field; the ratio between these values is off by a factor of 1.11. Hence, the gas conditions in the disk are not quite the same, and we assume that this should not affect the observed powers of the jets because the jet powers are estimated based on the measured radio-core luminosity. Consequently, our estimated mass accretion rate should be viewed in reference to our model assumptions, which is a constant accreted value along the flow all the way to the horizon. 

The procedure to constrain the mass accretion rate and the jet launching radius are as follow: (1) Initially, we use the Bondi mass accretion rate and the observed jets power values from A06 as our input parameter for our $\dot M$ and $\Qjet$. Keep in mind that our inferred $\dot M$ is a constant accreted value along the flow and through the event horizon. (2) We run these values through our model and search for $\epsilon_{_{-}}$, $\epsilon_{_{+}}$, $\ell$, $r_*$ that would reproduce the observed jets power at the shock location by calculating $P_{\rm disk} = \dot{M} \triangle \epsilon$. (3) Once we obtain a disk-structure that produces the observed jets power ($P_{\rm disk} = P_{\rm jet}$), we use the calculated dynamical gas flow speed $v(r)$ and run it through our transport equations to obtain the particle energy escaped rate $\Qesc$. The particle energy escape rate is constrained by the escaped parameter $A_0$, and consequently, the diffusion coefficient $\kappa_0$ as we have discussed at the end of Section~2.2. One or two $\kappa_0$ roots are possible for a particular flow. We search for a single-value of $\kappa_0$ that satisfies the condition in Equation (\ref{eq2a}) that also reproduces the escaped power from the disk using Equation (\ref{eq1c}). The single-root of $\kappa_0$ represents the smallest possible values of $\triangle \epsilon$ or the largest possible $\dot{M}$ that still yield the observed jet power for a selected source. Hence, in our model, $\dot{M}$ represents the upper limit mass accretion rate at the event horizon. The searching for the diffusion coefficient $\kappa_0$ is fully discussed in Figures (3a) and (3d) in Section 7.2 of LB05. If we obtain the $\Qesc$ value that is equal to the power escaped rate from the disk, then $\dot{M}$ is constrained. Otherwise,  we go back to step (1), modify $\dot{M}$, and go through steps 2 and 3 gain.

\subsection{$\rjet$, $\dot{M}$, $\Qjet$ Correlations}
Using the observed estimated jet powers for the sources from A06, our model satisfy both conditions in Equation (\ref{eq1b}), while constraining the mass accretion rate $\Mdot$ and the jet launching radius $\rjet$. The estimated values for these sources are in Table~\ref{tbl-1}. It is interesting to note that our inferred mass accretions are either similar or about a factor of 2 or 3 larger than the Bondi mass accretion rates as estimated by A06 for the selected sources. It is expected that the mass accretion rate at the event horizon of an ADAF with outflows/jets is smaller than the Bondi mass accretion rate, because most of the accreted mass at the Bondi radius will convert to jets or wind outflows before reaching the event horizon~\citep[e.g.,][]{qua03,yqn03}. From our previous study, we have shown that less than $20\%$ of the relativistic particles are being converted into the jets/outflows at the shock location, and less than $0.1\%$ of the accreted masses are being converted to the jets/outflows \citep[e.g.][]{lb07,bdl11}. This suggests that most of the mass loss could be in the form of disk-wind outflows \citep[e.g.,][]{yua15,mer16}, where wind and jets have been known to coexist in radio-loud AGNs \citep[e.g.,][]{tom12,fuk14}. Hence, if our model also accounts the mass loss due to wind outflows, then we expect our estimated mass accretion rate $\dot{M}$ should be lower than the Bondi mass accretion rate. However, it is beyond  the scope of this paper to determine if most of the mass loss come from the disk-wind outflows. Below, we use these sources to construct some possible correlations among the jet location, the observed jet power, and the accretion power $P_{\rm acc}=\eta \dot{M}c^2$, with a general assume value $\eta = 0.1$ relating the efficiency for the conversion of accreted rest mass into energy within the disk. 
\begfig[t] \hskip-0.0in \epsscale{1.16} \plottwo{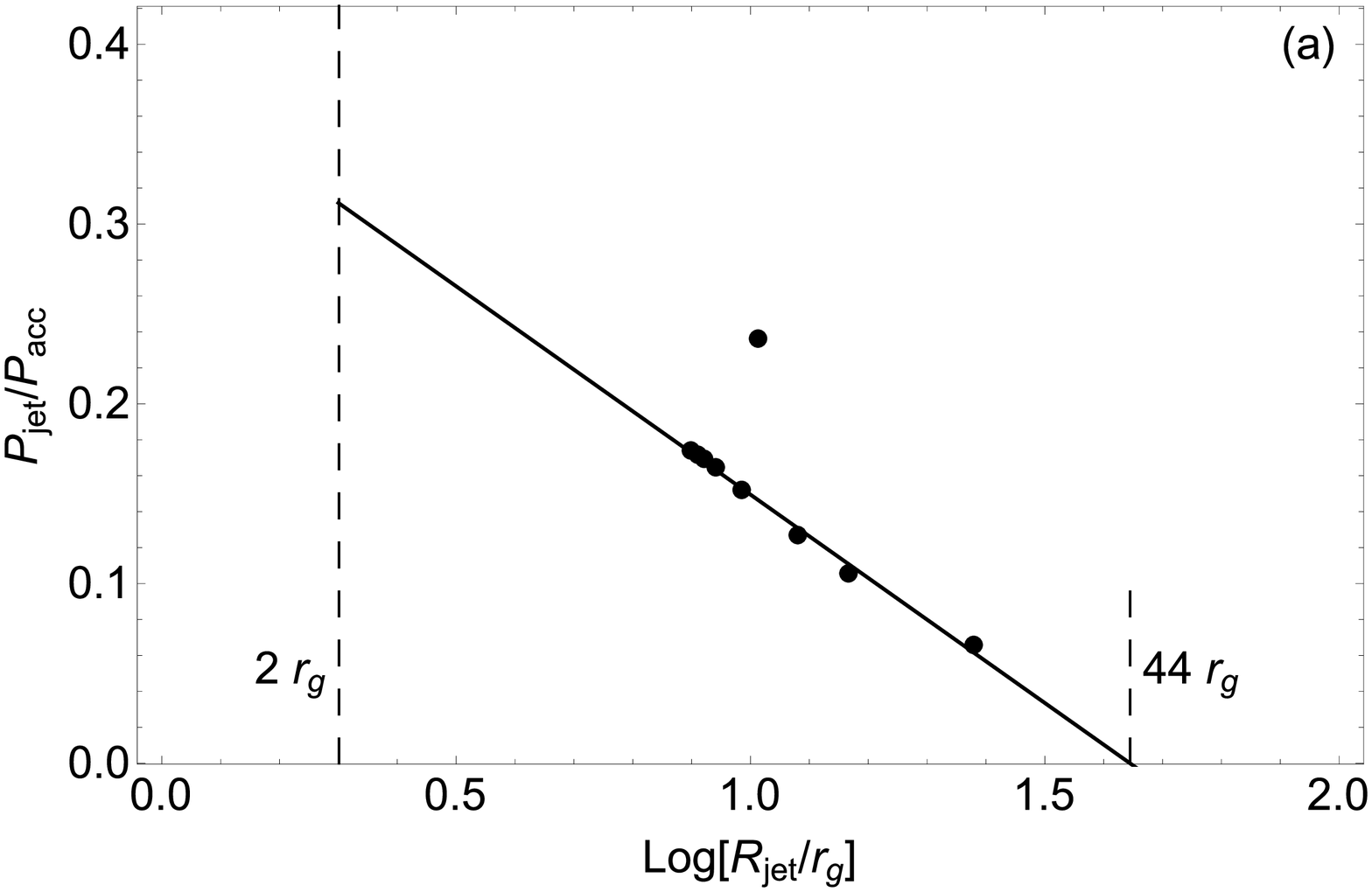}{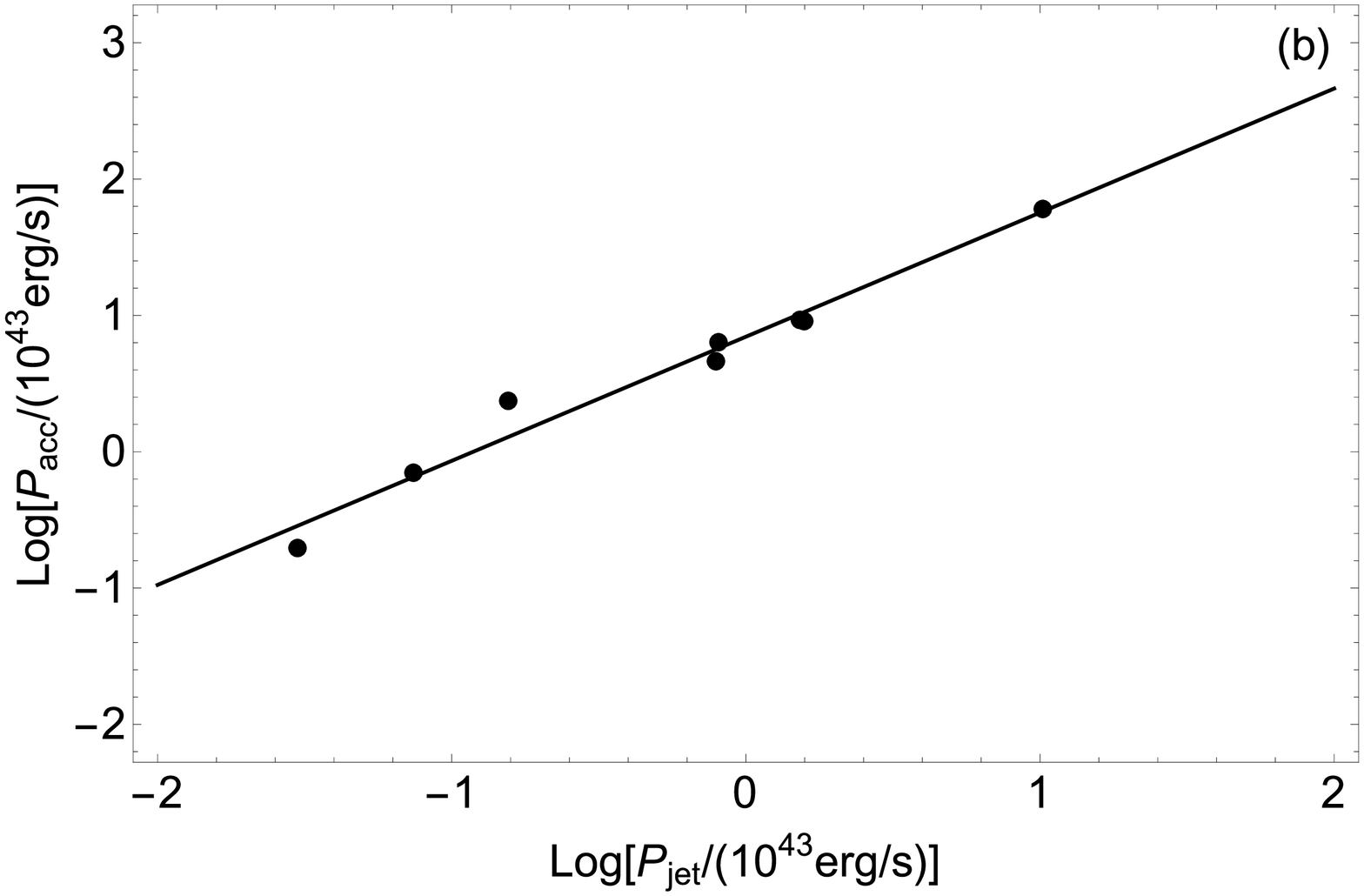}
\caption{\footnotesize (a) The energy conversion between the accreted power and the jet power at the associated estimated jet location. This plot shows that energy conversion is absent beyond $44 \, r_g$ and the jet power near the event horizon is about $30 \%$ of the accreted power. (b) The estimated radiated power at different observed estimated jet power. The solid circles and the solid lines are our estimated values and the best fits through the data, respectively. Plots (a) and (b) demonstrates a negative and positive correlations between the relative energy conversion and jet radius and the radiated power and the jet power, respectively. The outlier in plot (a) is for the source M87, and this data point is not included in our correlation studies.}
\label{fig4} %
\finfig

In Figure~(\ref{fig4}a) we plot the relative energy conversion between our calculated estimated accretion power to the observed estimated jet power at different inferred jet locations. The plot shows NGC 4486 (M87) as an outlier. After a careful examination, we notice that the Bondi mass accretion obtained by A06 is about an order of magnitude smaller than the value obtained by \citet{rey96} and \citet{kuo14}. Hence, this new value could potentially remove the outlier. For the purpose of using the same consistent data set, we decide not to use the Bondi value for M87 from the above authors, but instead, we ignore the data point NGC 4486 in the correlation study. The fit can be described by a mathematical expression of the form
\begeq 
\frac{\Qjet}{P_{\rm acc}} = A_1 + B_1 \, log \left(\frac{\rjet}{r_{g}}\right), 
\label{eq6a} 
\fineq
where $A_1 = 0.38$ and $B_1 = -0.23$ are our fitted values with a linear regression value of $R^2_{\rm value} = 0.99$, suggesting that this is a good fit. This correlation shows that a significant fraction of the accreted energy is required to convert the accreted mass to relativistic-energy particles for the production of the jets when the jets are launched near the horizon, and at least equal to about 15\% of the accreted energy is required around~10 $r_g$. However, only a fraction of the relativistic particles (about 20\%) will escape to form the jets/outflows, while the remaining relativistic particles will either get advected toward the horizon or get diffused outward radially away from the jets location, and only $0.1\%$ of the accreted masses are contributed to the outflows. Moreover, the result in Figure~(\ref{fig4}a) of our model also indicates that there is a jet/outflow cut-off or quenching at about $\sim 44 \, r_g$, suggesting that there is a minimum mass accretion rate or jet power where no jets/outflows can occur for low-power radio-loud AGNs. 

Most interestingly, \citet{nt15} addressed the energy efficiency of jet production by defining $\eta_{\rm jet} = P_{\rm jet}/\dot{M}_{\rm BH} c^2$, that provides the measurements of the jet power and the mass accretion rate onto the black hole for 27 low-luminosity active galactic nuclei (LLAGN) sources based on an ADAF model. Assuming the mass accretion at the event horizon is $\dot{M}\sim \alpha \dot{M}_B$, where $\alpha$ is the standard prescription for the viscosity \citep[e.g.,][]{ss73}, such that $\dot{M} \approx 30$ or 100\% of the Bondi mass accretion rate for $\alpha = 0.1$ and 1, respectively. They found that the distribution of jet efficiency is about 3 or 1\% for a disk with viscosity of 0.1 or 1, respectively. Their results seem to indicate that the jet efficiency would increase with decreasing viscosity. Our result on Figure (4a) is closely related to this jet efficiency factor, and it suggests a jet efficiency distribution of  6-18\% for an inviscid disk. More importantly, our result further suggests that the jet efficiency increases as the jet launching radius moves closer to the event horizon.

In Figure~(\ref{fig4}b), we plot the accreted power $P_{\rm acc}$ verses the jet power $P_{\rm jet}$ and find that there is a correlation between $\Qjet$ and $P_{\rm acc}$, which can be described by a power-law model of the form
\begeq 
log \left(\frac{P_{\rm acc}}{10^{43} \, \rm erg \,s^{-1}}\right) = A_2 + 
B_2 \, log \left(\frac{\Qjet}{10^{43} \,\rm erg \, s^{-1}}\right), 
\label{eq3a} 
\fineq
where $A_2 = 0.83$ and $B_2 = 0.91$ are our fitted values, and $R^2_{\rm value} = 0.97$ as the linear regression value. Interestingly, these fitted values are very similar to the values obtained by A06, and this correlation will allow us to estimate the mass accretion rates $\dot{M}$ for a given jet power of any low-power radio-loud sources. 
\begfig[t] \hskip-0.0in \epsscale{1.16} \plottwo{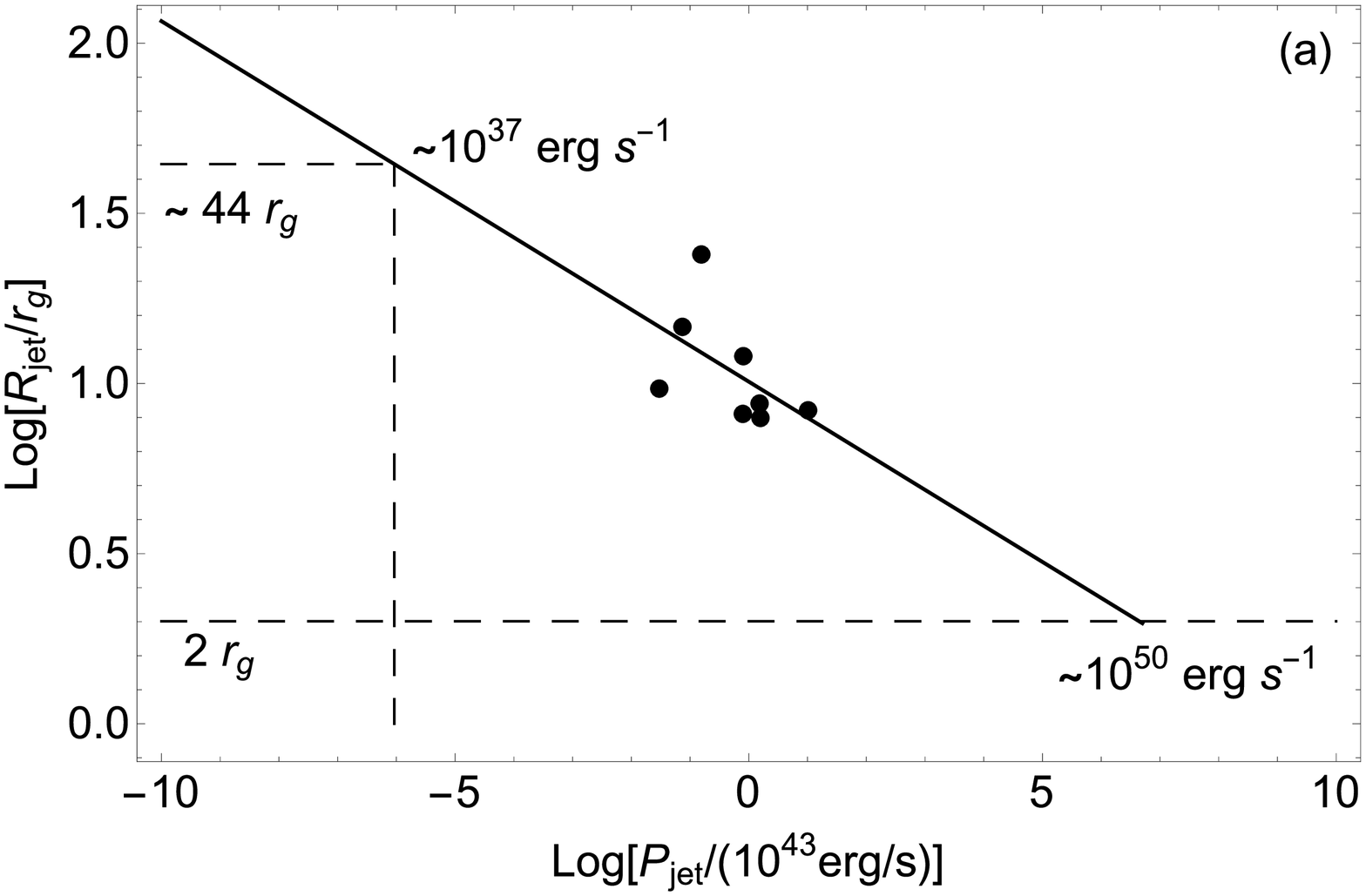}{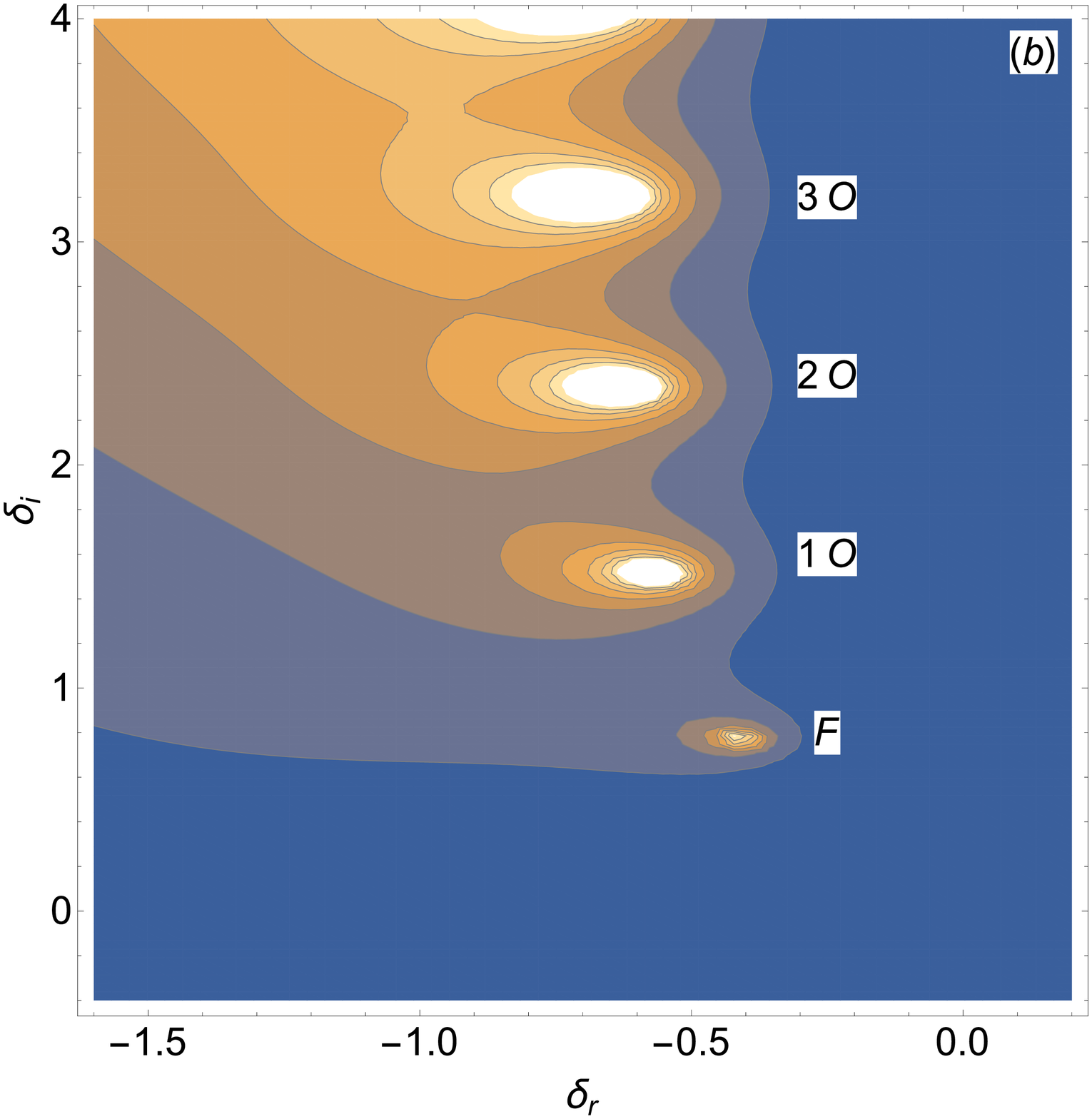}
\caption{\footnotesize (a) The estimated jet location for different observed estimated jet power. The solid circles and the solid lines are our estimated jet radii and the best fits through the data, respectively. We extrapolate the fit to $r = 2 \, r_g$ and $~44 \, r_g$, and our model indicates the jet power for low-power radio-loud AGNs cannot be less than $10^{37} \, {\rm erg \, s^{-1}}$ and more than $10^{50} \, {\rm erg \, s^{-1}}$, respectively.  (b) For illustration, we show the eigenfrequencies of the F-Mode and overtones for NGC 507. The results indicate the F-Mode and overtones are stable.} 
\label{fig5} %
\finfig
In Figure~(\ref{fig5}a), we plot the jet launching radius verses the observed jets power. The best-fit attempt described by a power-law model of the form
\begeq 
log \left(\frac{\rjet}{r_{g}}\right)  = A_3 + B_3 \, log \left(\frac{\Qjet}{10^{43} \,\rm erg \, s^{-1}}\right) 
\label{eq5a} 
\fineq
yields $A_3 = 1.01$ and $B_3 = -0.11$ with $R^2_{\rm value} = 0.25$ as  the linear regression value. This data is more scattered and less correlated than Figures (4a) and (4b). The plot shows that as the jets power increases, the jets launching location moves inward radially toward the event horizon suggesting that there are limits to the power of the jets. It also indicates that there is no jet/outflow power that can be greater than $10^{50} \, {\rm erg \, s^{-1}}$ or less than $10^{37} \, {\rm erg \, s^{-1}}$ beyond $44 \, r_g$. However, these values could be off because the jet launching radius and the jet observed power are not strongly correlated. The increasing power is expected, since the shock compression increases as the shock location moves closer to the horizon according to the values in Table-\ref{tbl-2}, and any outflows/jets that occur near the horizon have to be powerful to overcome the strong gravitational force at that point. 

It is also interesting to point out that, from our model, $P_{\rm jet}/P_{\rm acc} \propto \triangle \epsilon/\eta c^2$ and the energy dissipation across the shock $\triangle \epsilon$ is closely related to the upstream flow parameters (i.e., the conserved quantities such as $\epsilon_{-}$ and $\ell$), which in turn determines the shock location $r_* \sim R_{\rm jet}$. Since it is known that there exists a  correlation between $r_*$ and ($\ell, \epsilon_{-}$) \citep[e.g.,][]{dbl09,ck16}, hence, $P_{\rm jet}/P_{\rm acc}$ is expected to correlate with $R_{\rm jet}$ in theory as obtained in Figure 4(a). Likewise, a similar argument is applied to explain Figure 4(b) qualitatively. One can see in both figures that the effect of mass-accretion rate $\dot{M}$ will drop out. It is somewhat different, however, in Figure 5(a), because $P_{\rm jet}$ by itself is dependent on $\triangle \epsilon$ as well as $\dot{M}$ where $\dot{M}$, while conserved along a flow, is related to a combination of ($r,H,\rho,v$). Hence, the cause of the scatter in Figure 5(a) could be attributed to the scatter in the obtained $\dot{M}$ (for an individual source), because in this case the effect of $\dot{M}$ does not drop out as a function of $r_*$. Hence, the correlations in Figures~(4a) and (4b) are model dependent.
 
Assuming a jet power of $\Qjet = 3.44 \times 10^{43} \, {\rm erg \, s^{-1}}$ for NGC 4486 (M87) as indicated in Table~\ref{tbl-1}, using Equations~(\ref{eq3a}) and (\ref{eq5a}), we estimate the mass accretion rate and the jet launching radius for M87 to be about $\sim 0.038 \, {\rm M_{\odot} \,  yr^{-1}}$ and $\sim 9 \, r_g$, respectively. Most recently, the mass accretion rate for M87 near the horizon has been estimated to be about $\sim 0.001 \, {\rm M_{\odot} \, yr^{-1}}$ as an upper limit \citep[e.g.,][]{fwl16} with an assumed jets power of about $\sim 8 \times 10^{42} \, {\rm erg \, s^{-1}}$ \citep[e.g.,][]{rus13}. Using a value of $\sim 8 \times 10^{42} \, {\rm erg \, s^{-1}}$, our correlations give the mass accretion rate and the jet launching radius to be about $\sim 0.01 \, {\rm M_{\odot} \, yr^{-1}}$ and $\sim 10 \, r_g$, respectively. For \SgrA, \citet{fb99} and \citet{ybw12} have suggested the jets or outflows power to be between $10^{37}$ to  $5 \times 10^{38} \, {\rm erg \, s^{-1}}$. Adopting these values for \SgrA, our correlation functions give the mass accretion rate and the jet launching radius to be about $\sim 4.3 \times 10^{-5}$ to $1.5 \times 10^{-3} \, {\rm M_{\odot} \, yr^{-1}}$ and $1.2 \times 10^{-3}$ to $8.0 \times 10^{-4} \, r_g$, respectively. \citet{qnr99}, for example, have suggested that the Bondi mass accretion rate for \SgrA~to be $\sim 10^{-5} \, {\rm M_{\odot} \, yr^{-1}}$ with the jets launching radius of $3.4 \, r_g$ \citep[e.g.,][]{ymf02}. \citet{yqn03} and \citet{yus15} have also suggested the mass accretion rate for \SgrA~near the event horizon to be about $10^{-7} \, {\rm M_{\odot} \, yr^{-1}}$. Clearly, our estimated mass accretion rates at the event horizon for both M87 and \SgrA~are too large. Most interestingly,~\citet{wal16} and \citet{had16} have recently estimated the jets location of M87 to be within 5 -- 10 gravitational radii of the radio core at 43GHz and 86 GHz using the Very Long Baseline Array, respectively, which is consistent with our finding. However, it is interesting to note that the jet launching radius for \SgrA~is very close to the event horizon according to our model. This could be an indication that our model has reached its limit.  

It is important to mention again that, in the presence work, we only consider energy dissipation ($\triangle \epsilon$) only across a shock. In reality, such a dynamical process is expected to be accompanied by other losses in mass and angular momentum of accreting gas, all of which are carried away in the form of outflows (winds/jets) \citep[e.g.,][]{fk07}. For example, \citet{fk07} discussed general relativistic (GR) Rankine-Hugoniot shocks in hydrodynamical accretion flows in Kerr geometry taking account energy, mass, and angular momentum dissipation of gas across the shock. They found that mass loss rate can be quite significant with respect to accreting gas. We speculate that if mass or angular momentum dissipation of gas across the shock is included in our model, then the shock location would either move further outward away from the horizon or move further inward toward the horizon, respectively. We think that by removing mass or angular momentum of the gas at the shock location, this will relocate the centrifugal barrier to a higher/lower orbit causing the shock wave to form much further out/in, respectively.

Moreover, in our model, we assume a Schwarzschild's black hole and a pseudo-Newtonian potential to give the effect of GR. However, the jets launching radius will be larger, if GR or BH spin is taken into account. We speculate that when GR or BH spin is utilized, this would increase the accreted angular momentum due to frame-dragging \citep[e.g.,][]{kc17}. As a result, this will force the centrifugal barrier to be at a larger orbit causing a shock radius or the jet launching radius to move further away from the event horizon. Additionally, if global magnetic fields are also included in the calculation \citep[e.g.,][]{tt10}, we believe the gas is most likely be ionized and frozen-in to a large-scale magnetic field. However, it is not clear to us if this will cause the shock location to shift inward or outward without performing the actual calculation, and it is beyond the scope of this paper to perform such a calculation. Hence, if global magnetic field, black hole spin, and mass and angular momentum dissipation are included in the calculation, then depending on which of these components is dominate, we speculate a shock location will move in favor of that component.

The results for M87 and \SgrA~can be further tested against observations by measuring the average energy in this region of the disk. From our previous calculation, the mean energy of the relativistic particles in the disk at the shock location for M87 and \SgrA~are about $0.01$ ergs (e.g., LB05). According to observations, the observed brightness temperature near the base of the jets is between $10^{10} - 10^{13}$ K, however, the typical observed brightness temperatures are about $10^{11}$K \citep[e.g.,][]{kel07,kim16,sob17}; this gives about $2.1 \times 10^{-5}$ ergs as the mean energy of the high-energy particles near the jets launching radius. Our value is about 500 larger, and this clearly implies that first-order Fermi acceleration process is very efficient in accelerating the high-energy particles to power the jets/outflows. Because our value is too large, this suggests that we need to relax our way of constraining the diffusion coefficient $\kappa_0$ in the transport equation as we have discussed above, or allowing the outflows to occur downstream from the shock location as suggested by \citet{cha99}, and more recently by \citet{kc17}. Additionally, the enhancement of emission from radiation due to the presence of a shock is expected to be about a factor of $\sim 5-6$ comparing against the background gas when a shock is absent in the flow, and this is because we have shown that the mean energy of the relativistic particles in the disk is boosted by a factor of $\sim 5-6$ at the shock location (e.g., LB05). More importantly, the terminal (asymptotic) Lorentz factor of the jet $\Gamma_{\infty}$ predicts by our model for these sources is between $\sim 5-8$ (See Section 7.4 from LB05 for more details). \citet{ups91} predict that FR I radio galaxies should have jets with bulk flow speeds in the range from $\Gamma_\infty \sim 5$ to $\sim 35$, with most near $\Gamma_\infty \sim 7$, which is in good agreement with our values.

Additionally, the correlations in Figure~(\ref{fig4}b) and less so in Figure (\ref{fig5}a) further suggest that with higher mass accretion, the gas flow speed takes longer distance to reach supersonic speed before a shock can form. This indicates that the shock location moves inward toward the horizon as the mass accretion increases. Consequently, the shock location will cease to exist for large mass accretion rate, and hence, the absence of jets/outflows, and this could possibly suggesting that the disk structure is transitioning from low/hard state to high/soft state \citep[e.g.,][]{rm06,tch15}.  The results in Table~\ref{tbl-2} also indicate that a much stronger shock (with larger Mach number and compression ratio) can be formed when a shock location is closer to the event horizon, and other researchers found similar trend in their work~\citep[e.g.,][]{cha89,ly97,ly98,ft04,fk07}. This means large Mach number or compression ratio could allow a much more powerful jets to form. Hence, the analysis seem to suggest that powerful jets launch at radii closer to the event horizon. However, this trend may not necessary be true because it depends on the nature of the shock location. For example, \citet{ly97,ly98}, \citet{fk07}, and \citet{le16} showed that compression is decreasing with decreasing shock radius for the inner shock, while increasing with decreasing shock radius for the outer shock, and the jet launching radii in Table-2 are based on the inner shock solutions. We did not utilize the outer shock solutions because they fail to provide the power budget to support the observed jets power for the indicated sources.  Moreover, from Table~\ref{tbl-2}, the energy dissipation fraction across a shock, $\triangle \epsilon/\epsilon_{-}$, is about a factor of $\sim 4$ to 120. According to our current model, all unbounded energies are released at the shock point. However, in actuality, the unbounded energies can be released in a range of distance below a shock location as we have suggested earlier. For example, \citet{cha99} and \citet{kc17} allowed the outflows to occur between the shock location and the inner sonic point. We speculate the huge energy dissipation fraction can be reduced by allowing the outflows to occur in the downstream region of the shock or over a  small range of values in the vicinity of a shock \citep[e.g.,][]{leeb17}, since the particle acceleration process is strongest at the shock location.

To further test our correlation relations, we use sources from BBC08, who have also utilized a similar approach as A06 to estimate the jets power and Bondi mass accretion rate as we have discussed above. Using the observed estimated jets power obtained by BBC08, interestingly, our correlation Equation (\ref{eq3a}) requires the mass accretion rates for these sources to be much lower than the values suggested by BBC08 in comparison with A06, with an exception of NGC 4696 and IC 1459 as indicated in Table~\ref{tbl-3}. \citet{dcf01}, \citet{loe01}, \citet{gsb03}, and \citet{pel03} have, independently, estimated the Bondi mass accretions for NGC 1399, 3C 270, and IC 4296 to be about 0.04-1.0, 0.04, and 0.02 ${\rm M_{\odot} \, yr^{-1}}$, respectively, and we notice that these values are about an order of magnitude larger than our estimated $\dot{M}$ values. Currently, the Bondi mass accretion rates for other sources have not been independently measured by other researchers. 

It is clear that if outflows are presence, then the mass accretion rate cannot remain constant, but instead, it should decrease with decreasing radius. \citet{bb99} proposed a simple scaling accretion rate $\dot{M}(r) \equiv \dot{M}_{_{\rm in}} (r/r_{_{\rm in}})^{\beta}$, where $r_{_{\rm in}}$, $\dot{M}_{_{\rm in}}$, and $\beta$ are the innermost accretion radius, the mass accretion rate at this radius, and $\beta$ is a power-law index that measures the strength of the outflow; $\beta$ cannot exceed 1 for energetic reasons, while $\beta = 0$ corresponds to a constant mass accretion rate (no outflow). There are some observational evidences that $\beta \approx 0.5$ \citep[see][]{nt15}. If this mass accretion rate is utilized in our model, then we speculate that the gas density profile would decrease with radius causing a shock front to form much further away from the event horizon. This will force a shock point to be at a larger radius, consequently, the jet launching radius can be larger than $44 r_g$. Additionally, the mass accretion rate at the event horizon would be smaller than the values that we have estimated in Tables 1 and 3 for the indicated sources.

Finally, we analyze the stability of the shock locations using the \citet{le16} perturbation approach. We notice the fundamental mode and overtones of the disk structures that we model for the sources in A06 are stable.  This suggests that these sources can produce episodic or continuous jets/outflows. Additionally, when we explore the ($\epsilon$, $\ell$)-parameter space in the stable region, the Z-mode is increasingly stable but weak as we decrease $\ell$ while holding $\epsilon$ constant. Eventually, the stable Z-mode dissolves and changes to a stable, non-zero frequency mode. Our analyses show that the disk structures, based on our model, for all the sources in A06 contains no Z-mode. A demonstration of the stable fundamental and overtones with no Z-mode is shown in Figure~(\ref{fig5}b) for NGC 507. 

\section{CONCLUSIONS}

Using our theory for the production of the relativistic outflows observed from active galaxies, we estimate the mass accretion rate and jet launching radius associated with the observed estimated jets power for 52 low-power radio-loud AGNs from A06 and BBC08. Our results show a direct correlation between the calculated estimated radiation power and the observed estimated jets power that is consistent with the result obtained by A06. Furthermore, our model suggests that the Bondi mass accretion obtained by BBC08 for their sources are relatively high comparing to our estimated values. Additionally, our estimated mass accretions for NGC 1399, 3C 270, and IC 4296 from the correlation in Equation~(\ref{eq3a}) are similar with the Bondi values estimated by other researchers \citep[e.g.,][]{dcf01,loe01,gsb03,pel03}. Using our correlation relations, we estimate the mass accretion rates for M87 and \SgrA~to be about $\sim 0.01$ to $0.038 \, {\rm M_{\odot} \, yr^{-1}}$ and $\sim 4.3 \times 10^{-5}$ to $1.5 \times 10^{-3} \, {\rm M_{\odot} \, yr^{-1}}$ with the jets launching radii at less than $\sim 10 \, r_g$ and $\sim 1.2 \times 10^{-3} \, r_g$, respectively. The launching radius for M87 is within the observed estimated jet launching radius of a radio core at 86 GHz using the Very Long Baseline Array, but for \SgrA~our value is much smaller than the value estimated by \citet{ymf02}. Our estimated mass accretion rates $\dot M$ are about a factor of 2 higher than the Bondi values for the sources from A06, and are also about a magnitude higher than the observed estimated mass accretion rate near the event horizon for M87 and \SgrA. Moreover, the mean energy for the high-energy particles at the jets launching radius from our model is about 500 times larger than the observed mean energy near the base of the jets. We believe these disagreement can be improved by relaxing our approach in searching for the diffusion coefficient or allowing the launching point to occur downstream from the current shock location. In future work, we plan to incorporate the mass and angular momentum loss rates and the changing mass accretion rate due to disk-wind/jets outflows to our current model as well as to our working viscous model; we expect this will reduce the calculated estimated mass accretion rate at the event horizon while pushing the jet launching radius away from the event horizon. Finally, for the first time, our current model shows a strong negative correlation between the jets launching radius and the jets power. It indicates that powerful jets occur at small radii near the event horizon with jets/outflows cut-off or quenching beyond 44 $r_g$.

\acknowledgements T.L. would like to thank Peter Becker, Mario Gliozzi, George Chartas, Keigo Fukumura, Govind Menon, and Chris Done for useful discussions and comments, and also to the anonymous referee for the constructed comments and suggestions that significantly improved the content of the paper.  

{}

\newpage
\begin{deluxetable}{lccccc}
\tabletypesize{\small} %
\tablecaption{Jet Launching Radius for AGNs from \citet{all06}
\label{tbl-1}} %
\tablewidth{-0.pt} %
\tablehead{ %
\colhead{$\rm Name$} %
&\colhead{$M_{\rm BH}(10^{9}) \atop{(M_{\odot})^a}$}%
&\colhead{$P_{\rm jet}(10^{43})   \atop{(erg/s)^a}$} %
&\colhead{$\dot{M}_{\rm B}(10^{-3}) \atop{(M_{\odot}/yr)^a}$} %
&\colhead{$\dot{M}(10^{-3}) \atop{(M_{\odot}/yr)^b}$} %
&\colhead{$\rjet \atop{(r_g)^b}$} 
} \startdata
NGC 507         %
& 0.842 %
& 10.22        %
& 45.71    %
& 106.5 %
& 8.34 \\   %
NGC 4374         %
& 0.664 %
& 1.526        %
& 8.611    %
& 16.36 %
& 8.73  \\   %
NGC 4472         %
& 0.778 %
& 0.807        %
& 10.96   %
& 11.22    %
& 12.03  \\  %
NGC 4486   %
& 2.717 %
& 3.440        %
& 25.70    %
& 25.70 %
& 10.30  \\  %
NGC 4552         %
& 0.365 %
& 0.156        %
& 4.169    %
& 4.169 %
& 23.95  \\  %
NGC 4636         %
& 0.160 %
& 0.030        %
& 0.347    %
& 0.347 %
& 9.66   \\   %
NGC 4696         %
& 0.402 %
& 0.791        %
& 4.519    %
& 8.134 %
& 8.14  \\   %
NGC 5846         %
& 0.395 %
& 0.074        %
& 1.303    %
& 1.238 %
& 14.67  \\   %
NGC 6166         %
& 0.902 %
& 1.579        %
& 5.888    %
& 16.015 %
& 7.92  
\enddata

\tablecomments{$^a$ These are the observed estimated values obtained from~\citet{all06} paper. The above sources are used to construct the correlations, except, NGC 4486 (M87). $^b$ These are the calculated estimated values from this paper.}
\end{deluxetable}
\begin{deluxetable}{lcccccccc}
\tabletypesize{\small} %
\tablecaption{Model Values for AGNs from \citet{all06}} \label{tbl-2} %
\tablewidth{-0.pt} %
\tablehead{ %
\colhead{$\rm Name$} %
&\colhead{$\ell_0$} %
&\colhead{$\epsilon_{-} $} %
&\colhead{$\epsilon_{+} $} %
&\colhead{$r_{c}^{out} $} %
&\colhead{$r_{*}$} %
&\colhead{$r_{c}^{in} $} %
&\colhead{$\mathcal{M}_{-} $} %
&\colhead{${\cal R}_{_{*}}$}
} \startdata
NGC 507%
& 3.28827%
& 0.0003701%
& -0.0165692%
& 510.22%
& 8.34%
& 5.46%
& 1.21%
& 2.18%
\\
NGC 4373%
& 3.2704  %
& 0.0004304 %
& -0.016045 %
& 434.70 %
& 8.73 %
& 5.51%
& 1.20%
& 2.17%
\\
NGC 4472 %
& 3.1867 %
& 0.00089 %
& -0.0118063 %
& 194.78 %
& 12.03 %
& 5.71 %
& 1.17 %
& 2.05 %
\\
NGC 4486%
& 3.426%
& 0.0002%
& -0.0234263%
& 967.12%
& 10.30%
& 5.16%
& 1.59%
& 3.81%
\\
NGC 4552%
& 3.1318%
& 0.0015835%
& -0.00500687%
& 93.42%
& 23.94%
& 5.63%
& 1.12%
& 1.88%
\\
NGC 4636%
& 3.236%
& 0.000576%
& -0.0146553%
& 317.47%
& 9.66%
& 5.61%
& 1.19%
& 2.12%
\\
NGC 4696%
& 3.2978%
& 0.00034%
& -0.0168184%
& 557.96%
& 8.14%
& 5.43%
& 1.21%
& 2.19%
\\
NGC 5846%
& 3.1585%
& 0.0011564%
& -0.00941067%
& 142.35%
& 14.67%
& 5.73%
& 1.15%
& 1.99%
\\
NGC 6166%
& 3.3105%
& 0.000305%
& -0.0171076%
& 625.29%
& 7.92%
& 5.38%
& 1.21%
& 2.20 %
\enddata
\tablecomments {All quantities are expressed in gravitational units (G=c=1), and note that the shock radius and sonic points are normalized to $M$, where $M$ is the mass of a supermassive black hole for individual source. These are the model values for sources from~\citet{all06} paper.\\}
\end{deluxetable}

\newpage
\begin{deluxetable}{lccccc}
\tabletypesize{\small} %
\tablecaption{Jet Launching Radius for AGNs from \citet{bbc08}
\label{tbl-3}} %
\tablewidth{-0.pt} %
\tablehead{ %
\colhead{$\rm Name$} %
&\colhead{$M_{BH}(10^{9}) \atop{(M_{\odot})^a}$}%
&\colhead{$P_{\rm jet}(10^{43})   \atop{(erg/s)^a}$} %
&\colhead{$\dot{M}_B(10^{-3}) \atop{(M_{\odot}/yr)^a}$} %
&\colhead{$\dot{M}(10^{-3}) \atop{(M_{\odot}/yr)^b}$} %
&\colhead{$\rjet \atop{(r_g)^b}$} 
} \startdata
3C 028
& 0.98
& 3.3
& --
& 36.5
& 8.92
\\
3C 031
& 0.50
& 7.9
& --
& 80.8
& 8.13
\\
3C 066B        %
& $2.09$ %
& $17.8  $      %
& $350 $   %
& $169$ %
& $7.46 $   %
\\ 
3C 075 
& 0.69
& 6.7
& --
& 69.5
& 8.27
\\
3C 078
& 0.41
& 87.7
& --
& 722
& 6.30
\\
3C 083.1
& 3.24
& 4.9
& 2100
& 52.3
& 8.55
\\
3C084 
& 0.38
& 637.0
& --
& 4391
& 5.11
\\
3C 189
& 1.70
& 50.5
& --
& 437
& 6.68
\\
3C 264
&0.47
& 18.1
& --
& 172
& 7.45
\\
3C 270 %
& $0.52$%
& $5.9   $   %
& $460 $   %
& $61.9 $   %
& $8.39$   %
\\
3C 272.1
& 1.00
& 1.5
& 85
& 17.8
& 9.70
\\
3C 274
& 3.39
& 3.4
& 240
& 37.5
& 8.89
\\
3C 293
& 0.14
& 33.6
& --
& 302
& 6.98
\\
3C 296
& 0.63
& 11.3
& 4000
& 112
& 7.83
\\
3C 317
& 0.18
& 59.4
& --
& 507
& 6.57
\\
3C 338
& 0.78
& 1.6
& 55
& 18.9
& 9.63
\\
3C 346
& 1.38
& 354.9
& --
& 2580
& 5.43
\\
3C 348
&1.55
&37.7
& --
& 335
& 6.89
\\
3C 438
& 2.19
& 133.8
& --
& 1060
& 6.03
\\
3C 442
& 0.25
& 1.0
& --
& 12.3
& 10.1
\\
3C 449
& 0.35
& 4.6
& 530
&49.4
& 8.61
\\
3C  465
& 1.38
& 38.3
& 2500
& 340
& 6.88 
\\ \hline
UGC 0968
& 0.35
& 0.15
& --
& 2.19
& 12.4
\\
UGC 5902
& 0.10
& 0.024
& --
& 0.41
& 15.0
\\
UGC 6297
& 0.21
& 0.067
& 21
& 1.05
& 13.5
\\
UGC 7203
& 0.10
& 0.33
& --
& 4.49
& 11.4
\\
UGC 7386
& 0.27
& 1.5
& 160
& 17.8
& 9.70
\\
UGC 7629
& 0.60
&0.81
& 110
& 10.2
& 10.4
\\
UGC 7760
& 0.35
& 0.16
& 44
& 2.32
& 12.3
\\
UGC 7797
& 0.21
& 0.88
& --
& 11.0
& 10.3
\\
UGC 7878
& 0.14
& 0.03
& 47
& 0.51
& 14.7
\\
UGC 7898 %
& $2.00$%
& $0.34  $   %
& $110$   %
& $4.61 $   %
& $11.3 $   %
\\
UGC 8745 %
& $0.25$%
& $0.60  $   %
& $750 $   %
& $7.73 $   %
& $10.7 $   %
\\
UGC 9706
& 0.27
& 0.074
& 13
& 1.15
& 13.3
\\
UGC 9723
&0.05
& 0.14
& --
& 2.06
& 12.5
\\ \hline
NGC 0507
& $ 0.78 $
& $ 10.2 $
& $ 450 $
& $ 102 $
& $ 7.91 $
\\
NGC 1316
& 0.23
& 0.61
& --
& 7.85
& 10.7
\\
NGC 1399 %
& $1.17$%
& $0.22  $   %
& $490$   %
& $3.10$   %
& $11.9  $   %
\\
NGC 3557 %
& $0.47$%
& $0.74  $   %
& $190$   %
& $9.36$   %
& $10.5  $   %
\\
NGC 4696
& 0.35
& 0.79
& 3.4
& 9.93
& 10.4
\\
NGC 5128
& 0.24
& 4.5
& --
& 48.4
& 8.63
\\
NGC 5419
& 1.05
& 1.6
& --
& 18.9
& 9.63
\\ \hline
IC 1459%
& $1.51$%
& $7.7   $   %
& $53$   %
& $78.9$   %
& $8.15 $   %
\\
IC 4296 %
& $1.10$%
& $8.7    $   %
& $3000$   %
& $88.2$   %
& $8.05  $   %
\enddata

\tablecomments{$^a$These observed estimated values are obtained from~\citet{bbc08} paper.
$^b$ These calculated estimated values are from this paper.}
\end{deluxetable}

\end{document}